\documentclass{ws-ijmpa}

\begin{document}

\markboth{Kukulin, Pomerantsev}{Dibaryons as carriers of strong internucleon interactions}

\catchline{}{}{}{}{}

\title{ Dibaryons as carriers of strong internucleon interactions and a basis
for nuclear physics based on QCD}
\author {V.I. Kukulin and V.N. Pomerantsev}
\address{Institute of Nuclear Physics, Moscow State University, E-mail: kukulin@nucl-th.sinp.msu.ru}

\maketitle


\begin{abstract} New concept of intermediate- and short-range nuclear
force proposed by the authors a few years ago is discussed briefly. The general
concept is based on an assumption on generation of the dressed dibaryon in
intermediate state in $NN$ interaction. This new mechanism has been shown 
to lead not only to  numerous new predictions in hadronic physics but also should be
responsible for a large portion of nuclear binding energy and properties of
nuclear wavefunctions at high momenta.
\end{abstract}
\section{Why the dibaryon?}
In the talk we will try to convince the readers that dibaryons
should be considered as basic degrees of freedom in nuclear (similarly to nucleons and
pions) and also in hadronic physics. Moreover, they may serve as a most
appropriate basis for non-conventional nuclear physics based on QCD. These
claims are based on some strong arguments and also fundamental theory. Due to a
limitation of space, we present here a few such arguments only (for more
detailed arguments see e.g. refs.~\cite{1,2}):

(i) A few recent studies made independently~\cite{3} have demonstrated
with evidence that traditional OBE-based approach to the basic
intermediate-range attraction in $NN$ system, which uses the scalar
$\sigma$-meson exchange (or the $t$-channel two-pion exchange with intermediate
$\pi\pi$ interaction in scalar-isoscalar channel) fails to reproduce the strong
intermediate-range $NN$ attraction which is basis for nuclear binding in all
traditional force models.

(ii) The values of cut-off parameters $\Lambda_{mNN}$ and
$\Lambda_{mN\Delta}$ ($m=\pi,\;\rho$,\dots) taken in all OBE-based $NN$ models
($\Lambda_{mNN}\sim 1.5\div 1.7$~GeV/c) are at least in 2 --3 times higher than
the values needed to describe quantitatively the $\pi^{\pm}$-production in
$pp$-collisions or $\pi$-meson absorption in $d$, $^3$He etc. and also as
compared with all theoretical predictions~\cite{1,2}. So, there is no one
consistent choice for the cut-off parameters $\Lambda_x$ able to describe
correctly both elastic and inelastic $NN$ scattering or $\pi$-meson production
and absorption.

(iii) Many hadronic experiments with nuclei done in the kinematical
region forbidden for a single-nucleon interaction (the so-called cumulative
experiments) show a large yield of particle products, which can be explained
only by an interaction of incident hadron with tightly correlated few-nucleon
clusters in nuclei, the degree of short-range correlations being incompatible
with any traditional nuclear force models, but may be interpreted quite
naturally as a manifestation of multi-quark (i.e. $6q$-, $9q$-,\dots) bags in
nuclei. Moreover, it was claimed by Jaffe many years ago that QCD does not
forbid the existence of such multi-quark bags and if {\em they do not exist
there should be the special QCD-based forbidness which we do not know
now.} So, the incorporation of dibaryons in hadronic and nuclear physics might
help to overcome such fundamental difficulties and to put a proper
cornerstone to the ground of QCD-based nuclear physics.

\section{Microscopic and EFT approach$\!$ to$\!$ intermediate- and short-range $NN$
interaction based on intermediate dibaryon generation}
If the mass of mesons exchanged between two nucleons exceeds 500~MeV (this
relates to all mesons except pion) the respective characteristic scale for
meson-exchange $NN$ interaction is less than 0.5~fm, i.e. the meson exchange
proceeds when two nucleons overlap deeply and thus the picture of two isolated
nucleons which are exchanged with a such heavy meson gets fully meaningless. The picture
becomes much closer to a unified $6q$-bag surrounded with $\pi$-, $\sigma$,
$\rho$-, etc. mesonic fields. We developed the respective microscopic
quark-meson model for such dressed six-quark bag as an intermediate in a
short-range $NN$ interaction (see Fig.~1)~\cite{1,2}:

\mbox{ \includegraphics[width=0.6\textwidth,keepaspectratio]{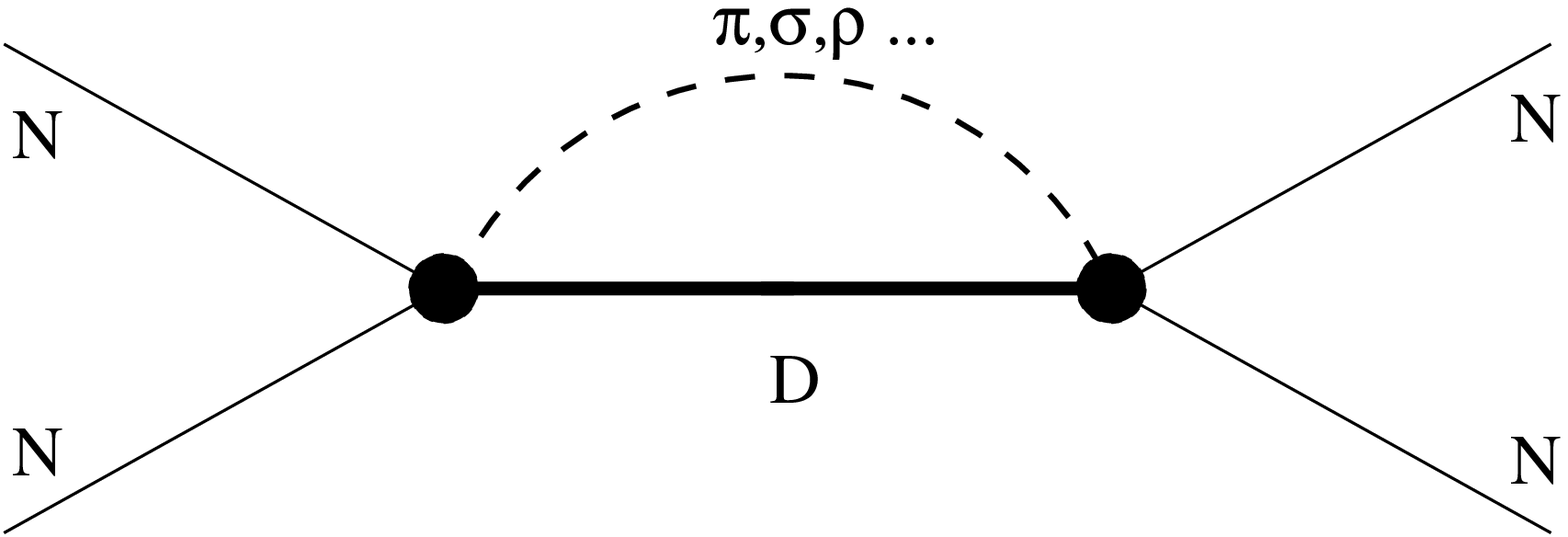}	
\qquad \parbox[b]{0.3\textwidth}{\small Fig.~1. $s$-channel mechanism for $NN$ interaction with an
intermediate dibaryon dressed with $\pi$-, $\sigma$,
$\rho$-, etc. mesonic fields.\hspace{\textwidth}{}}}

\noindent This $s$-channel mechanism replaces the conventional $t$-channel (i.e.
Yukawa-like) mesonic exchange at ranges $r_{NN}\lesssim 1\, {\rm fm}$. Using
this mechanism as a guide we constructed~\cite{2} the respective potential model
for $NN$ interaction which easily fits (with a few free parameters only) the $NN$
phase shifts at lower partial waves until 1~GeV and higher. This simple model
predicts the deuteron properties even more accurately than the best modern 
phenomenological $NN$ potentials like Argonne, Nijmegen etc. models.

Recently, this new force model has been reformulated~\cite{1} using a fully
covariant effective field theory (EFT) approach. In the approach the
intermediate dibaryon is described on the basis of $NN\to D(CC) \to NN$
transition ($CC$ means here two color three-quark clusters connected by a
gluonic string which form the dibaryon $D$). The amplitude of this transition
can be represented through the non-local Lagrangian density:
\[M_{fi}=i\langle 4,3|\mathrm{T}\left [e^{i\int
dxdx'\mathcal{L}_{\mathrm{int}}(x,x')}\right ] |2,1\rangle-
i\langle 4,3|\mathrm{I}|2,1\rangle, \]
where T means time-ordering while the bra and ket $|2,1\rangle$ and $\langle 4,3|$ relates to the
initial and final nucleons with 4-momenta $p_2,p_1$ and $p_4,p_3$ respectively. 
The nonlocal Lagrangian density 
$ \mathcal{L}_{\mathrm{int}}(x_1,x_2)=\mathcal{L}_{DN}+\mathcal{L}_{Dm}$,
includes two terms: $\mathcal{L}_{DN}$ describes the transition to a bare
dibaryon state with possible spins $S=0,1$ while $\mathcal{L}_{Dm}$ represents
the dressing of the bare $6q$-bag with its meson cloud~\cite{1}. The amplitude
$M_{fi}$ can be expressed through the dibaryon wavefunctions and respective
dibaryon propagators ${\cal G}^s(x_1,x_2;x_3,x_4)$ for which we solve the Dyson equation
using the basis of relativistic (Dirac) oscillator. After lengthy algebra one
gets eventually the analytical expression for relativistic $NN$ potential which
includes also an imaginary part responsible for the meson production in $NN$
collisions.
The potential includes the matrix elements of polarization operator, the latter
can be represented through the graphs:
\setcounter{figure}{1} 
\begin{figure}[h]
\centerline{
\includegraphics[width=0.75\textwidth,keepaspectratio]{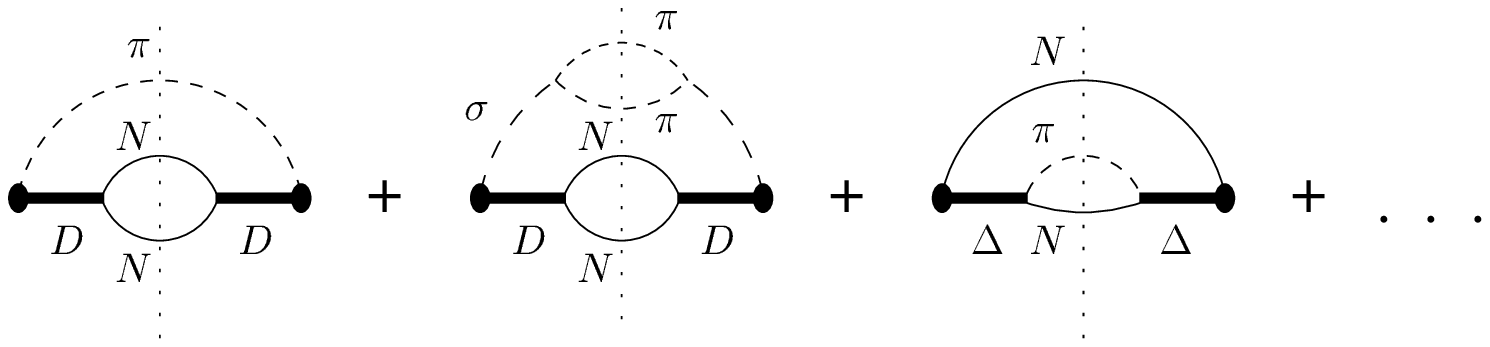}}
\caption{Several of the possible loops taken into account in calculation of polarization
operator of dibaryon. They correspond to the dressed bag state.\label{fig2}}
\end{figure}

\noindent If the total energy exceeds that of the $2\pi$-threshold the $2\pi$-production
process in scalar-isoscalar channel will proceed via intermediate (renormalized)
$\sigma$-meson with enhanced probability~\cite{1,2} which can naturally explain a
few puzzles related to $2\pi$-production cross sections in $pp$, $pn$, $pd$ etc.
collisions (e.g. the ABC puzzle etc.).

\section{Dibaryons in nuclei and hadronic processes}
Appearance of dibaryon mode in the fundamental $NN$ interaction must result in
the appearance of the dibaryon components in nuclear wavefunctions. In fact, it
was shown in our extensive $3N$ calculations~\cite{4} with the dibaryon model
for nuclear force that the presence of the strong scalar-isoscalar field in the
dibaryon leads to the strong exchange force between dibaryon and nucleons
surrounding it (this force is nothing else but a specific three-body force (see
Fig.~3). It is important to stress here that the $\sigma$-meson mass and width in
these graphs are renormalized noticeably due to partial chiral symmetry
restoration around the dense multiquark bag.
\begin{figure}[h]
\centerline{
\includegraphics[width=0.75\textwidth,keepaspectratio]{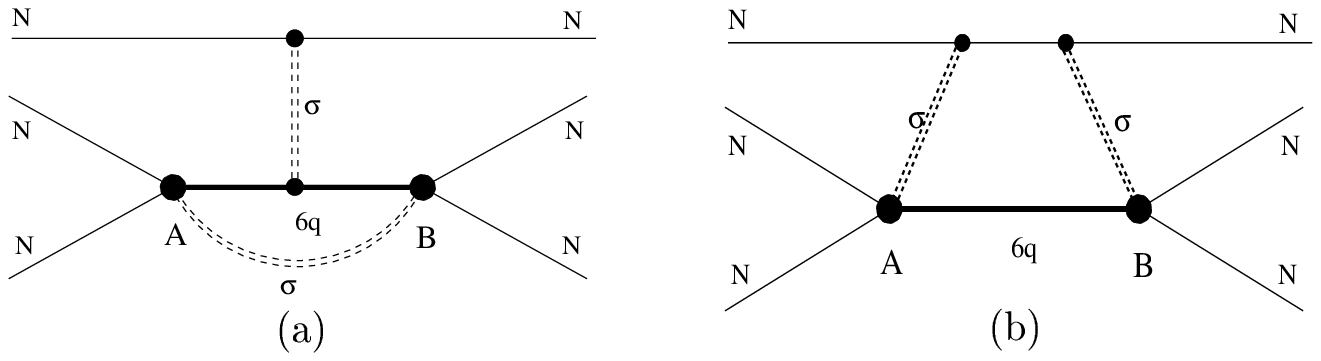}}
\caption{The graphs illustrating the new $3N$ scalar force induced by $\sigma$-exchange between
the dressed bag and third nucleon.\label{fig3}}
\end{figure}
So, the $\sigma$-meson mass gets much lower and has been estimated to be the
value $m_{\sigma}\sim 350\div 380$~MeV which should be compared to the free
$\sigma$-meson mass $\sim 550$~MeV. So that, the renormalized $\sigma$-meson
around the bag resembles the quasi-stable scalar $\sigma$-meson of the old OBE
models. The exchange by the quasi-stable $\sigma$-meson leads inevitably to
strong attractive force between dibaryon and other nucleons.In our $3N$
calculations~\cite{4} we have found that this new three-body force gives, at
least, a half of the total nuclear binding energy and contributes strongly to
other important observables. Moreover, the total weight of the dibaryon component
in $3N$ wavefunction is as large as 10\% 
or even higher. This large admixture of the very compact multi-quark bag
components in all nuclei should play decisive role in short-range $NN$
correlations and description of nuclear properties at high momentum and energy
transfers.

Furthermore, according to the general principles of quantum theory a new degree of
freedom must lead to respective new currents, e.g. in e.-m. processes. So that we
derived such a new isoscalar current in deuteron, which gives very essential M1-
and E2-contributions to circular polarization of $\gamma$-quanta in $\vec{n}p\to 
d\vec{\gamma}$ radiation capture process with thermal neutrons~\cite{5}, and also
to deuteron magnetic form factor at $Q^2\sim 1$~GeV and some corrections to
deuteron magnetic moment.
The intermediate dibaryon should contribute strongly also to numerous hadronic
processes like single- and multi-meson production and near-threshold
production of heavy mesons ($\rho$, $\omega$ \dots) in $pp$, $pd$ etc. collisions
at intermediate energies 1 - 5~GeV~\cite{1,2}, in numerous electro- and
photoproduction processes like $d(\gamma,2\pi^0)$ and two-nucleon electro- and
photo-disintegration processes like $^3{\rm He}(e,e'pp)$, $A(\gamma, pp)$ etc.

Thus, the characteristic features in the majority of hadron-nucleus
or photon-nucleus processes accompanying with high momentum or energy transfer must be
related to the interaction of high-energy projectile with dibaryon as
whole. The latter process should be described in terms of QCD-based approaches. 
Hence, one can summarize: {\em the dibaryon physics can be viewed as a very appropriate
``window'' through  which the full QCD enter the whole nuclear physics.}

The authors appreciate the financial support of their work from RFBR (grants
nos. 05-02-17404, 05-02-04000). V.I.K. thanks very much the Conference Organizers
for partial financial support of his participation in the Conference.

\end{document}